\documentclass{bmcart}
\usepackage{amsmath}
\usepackage[super,sort&compress,comma]{natbib}
\usepackage{hyperref}
\hypersetup{ colorlinks,
    citecolor=blue,
    filecolor=blue,
    linkcolor=blue,
    urlcolor=blue
}
\usepackage{graphicx}

\usepackage{subcaption}
\usepackage{array}
\newcolumntype{L}[1]{>{\raggedright\let\newline\\\arraybackslash\hspace{0pt}}m{#1}}
\newcolumntype{C}[1]{>{\centering\let\newline\\\arraybackslash\hspace{0pt}}m{#1}}
\newcolumntype{R}[1]{>{\raggedleft\let\newline\\\arraybackslash\hspace{0pt}}m{#1}} 
\usepackage[bmc]{bmcart-biblio}

\usepackage[polish,english]{babel}
\usepackage[utf8]{inputenc}
\usepackage[T1]{fontenc}
\usepackage{caption}
\begin{document}

%%% Start of article front matter
\begin{frontmatter}
\begin{fmbox}
\vspace{-1.cm}
\dochead{Research}
\vspace{-0.4cm}
\title{Estimating relationship between the Time Over Threshold and energy
loss by photons in plastic scintillators used in the J-PET scanner }

%%%%%%%%%%%%%%%%%%%%%%%%%%%%%%%%%%%%%%%%%%%%%%
%%                                          %%
%% Enter the authors here                   %%
%%                                          %%
%% Specify information, if available,       %%
%% in the form:                             %%
%%   <key>={<id1>,<id2>}                    %%
%%   <key>=                                 %%
%% Comment or delete the keys which are     %%
%% not used. Repeat \author command as much %%
%% as required.                             %%
%%                                          %%
%%%%%%%%%%%%%%%%%%%%%%%%%%%%%%%%%%%%%%%%%%%%%%

%author[
%   addressref={aff1},                   % id's of addresses, e.g. {aff1,aff2}
%   corref={aff1},                       % id of corresponding address, if any
%   noteref={n1},                        % id's of article notes, if any
%   email={jane.e.doe@cambridge.co.uk}   % email address
%]{\inits{JE}\fnm{Jane E} \snm{Doe}}
%
\vspace{-0.1cm}
\author[
   addressref={UJ},
   email={sushil.sharma@uj.edu.pl},
]{\fnm{S.}~\snm{Sharma}}
\author[
   addressref={UJ},
]{\fnm{J.}~\snm{Chhokar}}
\author[
   addressref={INFN},
]{\fnm{C.}~\snm{Curceanu}}
\author[
   addressref={UJ},
 ]{\fnm{E.}~\snm{Czerwi{\'n}ski}}
\author[
   addressref={UJ},
]{\fnm{M.}~\snm{Dadgar}}
\author[
   addressref={UJ},
]{\fnm{K.}~\snm{Dulski}}
\author[
   addressref={PAN},
]{\fnm{J.}~\snm{Gajewski}}
\author[
   addressref={UJ},
]{\fnm{A.}~\snm{Gajos}}
\author[
   addressref={UMCS},
]{\fnm{M.}~\snm{Gorgol}}
\author[
   addressref={UMCS},
]{\fnm{N.}~\snm{Gupta-Sharma}}
\author[
   addressref={INFN},
]{\fnm{R.}~\snm{Del~Grande}}
\author[
   addressref={UV},
]{\fnm{B.~C.}~\snm{Hiesmayr}}
\author[
   addressref={UMCS},
 ]{\fnm{B.}~\snm{Jasi{\'n}ska}}
\author[
   addressref={UJ},
]{\fnm{K.}~\snm{Kacprzak}}
\author[
   addressref={UJ},
 ]{\fnm{{\L}.}~\snm{Kap{\l}on}}
\author[
   addressref={UJ},
]{\fnm{H.}~\snm{Karimi}}
\author[
   addressref={UJ},
]{\fnm{D.}~\snm{Kisielewska}}
\author[
   addressref={NCBJ},
]{\fnm{K.}~\snm{Klimaszewski}}
\author[
   addressref={UJ},
]{\fnm{G.}~\snm{Korcyl}}
\author[
   addressref={NCBJ},
]{\fnm{P.}~\snm{Kowalski}}
\author[
   addressref={UJ},
]{\fnm{T.}~\snm{Kozik}}
\author[
   addressref={UJ},
]{\fnm{N.}~\snm{Krawczyk}}
\author[
   addressref={NCBJ-HE},
 ]{\fnm{W.}~\snm{Krzemie{\'n}}}
\author[
   addressref={UJ},
]{\fnm{E.}~\snm{Kubicz}}
\author[
   addressref={UJ,Mosul},
]{\fnm{M.}~\snm{Mohammed}}
\author[
   addressref={UJ},
 ]{\fnm{Sz.}~\snm{Nied{\'z}wiecki}}
\author[
   addressref={UJ},
 ]{\fnm{M.}~\snm{Pa{\l}ka}}
\author[
   addressref={UJ},
 ]{\fnm{M.}~\snm{Pawlik-Nied{\'z}wiecka}}
\author[
   addressref={NCBJ},
 ]{\fnm{L.}~\snm{Raczy{\'n}ski}}
\author[
   addressref={UJ},
]{\fnm{J.}~\snm{Raj}}
\author[
   addressref={PAN},
 ]{\fnm{A.}~\snm{Ruci{\'n}ski}}
\author[
   addressref={UJ},
]{\snm{Shivani}}
\author[
   addressref={NCBJ},
]{\fnm{R.~Y.}~\snm{Shopa}}
\author[
   addressref={UJ},
]{\fnm{M.}~\snm{Silarski}}
\author[
   addressref={UJ,INFN},
]{\fnm{M.}~\snm{Skurzok}}
\author[
   addressref={UJ},
 ]{\fnm{E.~{\L}.}~\snm{St{\k{e}}pie\'n}}
\author[
   addressref={NCBJ-HE},
 ]{\fnm{W.}~\snm{Wi{\'s}licki}}
\author[
   addressref={UMCS},
 ]{\fnm{B.}~\snm{Zgardzi{\'n}ska}}
\author[
   addressref={UJ},
 ]{\fnm{P.}~\snm{Moskal}}
%
%%%%%%%%%%%%%%%%%%%%%%%%%%%%%%%%%%%%%%%%%%%%%%
%%                                          %%
%% Enter the authors' addresses here        %%
%%                                          %%
%% Repeat \address commands as much as      %%
%% required.                                %%
%%                                          %%
%%%%%%%%%%%%%%%%%%%%%%%%%%%%%%%%%%%%%%%%%%%%%%
%
\address[id=UJ]{%
  \orgname{Faculty of Physics, Astronomy and Applied Computer Science, Jagiellonian University}, 
  \street{prof. Stanis{\l}awa {\L}ojasiewicza 11},                
 \postcode{30-348}                                
  \city{Cracow},                              
  \cny{Poland}                                   
}
\address[id=INFN]{%
  \orgname{INFN, Laboratori Nazionali di Frascati},
  %\street{},
  \postcode{00044}
  \city{Frascati},
  \cny{Italy}
}
\address[id=PAN]{%
  \orgname{Institute of Nuclear Physics PAN},
  %\street{},
  %\postcode{}
  \city{Cracow},
  \cny{Poland}
}
\address[id=UMCS]{%
  \orgname{Institute of Physics, Maria Curie-Sk\l odowska University},
  %\street{},
  \postcode{20-031}
  \city{Lublin},
  \cny{Poland}
}
\address[id=UV]{%
  \orgname{Faculty of Physics, University of Vienna},
  %\street{},
  \postcode{1090}
  \city{Vienna},
  \cny{Austria}
}

\address[id=NCBJ]{%
  \orgname{Department of Complex Systems, National Centre for Nuclear Research},
  %\street{},
  \postcode{05-400}
  \city{Otwock-\'Swierk},
  \cny{Poland}
}
\address[id=NCBJ-HE]{%
  \orgname{High Energy Physics Division, National Centre for Nuclear Research},
  %\street{},
  \postcode{05-400}
  \city{Otwock-\'Swierk},
  \cny{Poland}
}
\address[id=UJCM]{%
  \orgname{2nd Department of General Surgery, 
  Jagiellonian University Medical College},
  %\street{Kopernika 21},
  %\postcode{31-501}
  \city{Cracow},
  \cny{Poland}
}
\address[id=Mosul]{%
  \orgname{Department of Physics, College of Education for Pure Sciences, University of Mosul},
  %\street{},
  %\postcode{}
  \city{Mosul},
  \cny{Iraq}
}
%\address[id=Ghent]{%
%  \orgname{Department of Electronics and Information Systems, MEDISIP, %Ghent University-IBiTech},
%  \street{De Pintelaan 185 block B},
%  \postcode{B-9000}
%  \city{Ghent},
%  \cny{Belgium}
%}

\begin{artnotes}
%\note{Sample of title note}     % note to the article
%\note[id=n1]{Equal contributor} % note, connected to author
\end{artnotes}

\end{fmbox}% comment this for two column layout
\begin{abstractbox}
\vspace{-0.2cm}
\begin{abstract}
  \parttitle{Purpose} 
  Time-Over-Threshold (TOT) technique is being used  widely due to its implications in developing the multi-channel readouts mainly when fast signal processing is required. Using TOT technique as a measure of energy loss instead of charge integration methods significantly reduces the signals readout cost by combining the time and energy information. Therefore, this approach can potentially be used in J-PET tomograph which is build from plastic scintillators characterized by fast light signals. The drawback in adopting this technique is lying in the non-linear correlation between input energy loss and TOT of the signal. The main motivation behind this work is to develop the relationship between TOT and energy loss and validate it with the J-PET tomograph. \\
  \parttitle{Methods} 
  The experiment was performed using the $^{22}$Na beta emitter source placed in the center of the J-PET tomograph. One can obtain primary photons of two different energies: 511 keV photon from the annihilation of positron (direct annihilation or through the formation of para-Positronim atom or pick-off process of ortho-Positronium atoms), and 1275 keV prompt photon. This allows to study the correlation between TOT values and energy loss for energy range up to 1000 keV. As the photon interacts dominantly via Compton scattering inside the plastic scintillator, there is no direct information of primary photon's energy. However, using the J-PET geometry one can measure the scattering angle of the interacting photon. Since, $^{22}$Na source emits photons of two different energies, it is required to know unambiguously the energy of incident photons and its corresponding scattering angle for the estimation of energy deposition. In this work, dedicated algorithms were developed to tag the photons of different energies and studying their scattering angles to calculate the energy deposition by the interacting photons.\\
  \parttitle{Results} A new method was elaborated to measure the energy loss by interacting photons with plastic scintillators used in J-PET tomograph. It is found that relationship between the energy loss and Time Over Threshold is non-linear and can be described by function TOT = A0 + A1 * ln(E$_{dep}$ + A2) + A3 * (ln(E$_{dep}$ + A2))$^2$. 
  \parttitle{Conclusions} 
  A relationship between Time Over Threshold and energy loss by interacting photons inside the plastic scintillators used in J-PET scanner is established for a energy deposited range 100-1000 keV. 

\end{abstract}

\end{abstractbox}

\end{frontmatter}

\section{Background}
\label{sec::introduction}
Time-Over-Threshold~(TOT) technique was introduced first time by Nygren and Millaud~\cite{Nygren1991} and proved to be an excellent solution for the multi-channel readouts~\cite{Kipnis1997}. In the TOT method, one basically measures the signal pulse width at the selected thresholds, which can be used as the estimate of the signal's charge. For the energy deposition estimation this method is less precise in comparison to charge integration method. However, it reduces the readout cost by using only time to digital converter (TDC) combining both timing and energy information. Application of TOT method for the energy loss determination may be of particular advantage in the newly developed J-PET positron emission tomograph~\cite{Moskal2016, Moskal2018} which is based on plastic scintillators characterized by fast light signals with rise and decay times of the order of $\approx$~1 ns~\cite{SAINTGOBAIN, Wieczorek2017} and thus being about two orders of magnitude shorter than signals from crystals used in the current PET devices~\cite{CIMA2006, RMAO, LABR3}. Therefore, application of the TOT method in case of the J-PET tomograph build from plastic scintillators will enable fast signal processing reducing significantly signal acquisition dead time with respect to the crystal based tomographs.  
Despite of advantages providing the compactness of signals readout and low power consumption, the TOT technique confronts the challenge in terms of non-linear input energy to pulse width conversion~\cite{Olcott2008,fujiwara2010,Orita2018}.~It has been reported that using the multiple fixed triggering thresholds~\cite{Kim2009,Grant2014} or dynamic threshold levels~\cite{Shimazoe2012,Orita2015} for estimating the TOT values alleviate the problem of non-linearity to significant extent.\newline
~In Positron Emission Tomograph (PET) applications, the precise determination of the photon's interaction position, hit time and energy loss is crucial. Currently in PET scanners, crystal scintillators are used and they enable to determine energy of interacting photons as they undergo photoelectric effect~\cite{VAN2016,Slomka2016}.~But in plastic scintillators the incident photon interacts via Compton scattering depositing only part of its energy.~So, there is no direct information of energy deposition of the interacting photon.\newline
TOT studies with the plastic scintillators are very scarce~\cite{Jin-Jie2008}.
~In the recent work of Ashrafi and Gol~\cite{Ashrafi2011}, energy calibration of a plastic scintillator based on the Compton scattering and observing the light yield by using various monochromatic photon source was reported.~In this work we take advantage of the multilayer cylindrical acceptance of the J-PET scanner~\cite{Moskal2016,Kowalski2016A} which allows to determine the direction of photon before and after the scattering and by that gives access to its scattering angle~\cite{Moskal2018,Kowalski2018A}.~Thus, knowing the incident energy and the measured scattering angle of initial photon, the deposited energy in the photon's interaction can be calculated~\cite{Compton1923}.\newline
For the present study, $^{22}$Na source was used which emits photons of two different energies 511 keV (annihilation) and 1275 keV (prompt).~The 511 keV photons are originating from the direct electron-positron annihilation, from the decay of para-positronium and from annihilation of ortho-positronium atoms via the pick-off processes~\cite{Garwin1953,Moskal2019, Moskal2019NatureReview}. Since the aim of the study is to establish the relationship between TOT and energy loss it is required to develop an algorithm to clearly distinguish between the annihilation and prompt photons emitted from the source as well as a method of the proper association of measured signals to initial and scattered photons. To achieve this, we performed studies bases on the assumptions that 511 keV photons emitted in the annihilation of e$^+$e$^-$ are always back-to-back due to the momentum conservation. Secondly, to identify the prompt photons, we take advantage of the decay of long-lived ortho-positronium atoms~\cite{Dulski2018} which in the XAD$-$4 porous polymer~\cite{Jasinska2016} on the average decay in about 90 ns after the emission of a prompt photon.~In this article we present a method for establishing a relationship between TOT and the energy deposition.
\section{Methods}
\label{sec::materials}
\begin{figure}[h]
    \centering
    \begin{subfigure}{.5\textwidth}{
%     \subfigure[\label{PhotoJPET}]{%
     \includegraphics*[width=\textwidth]{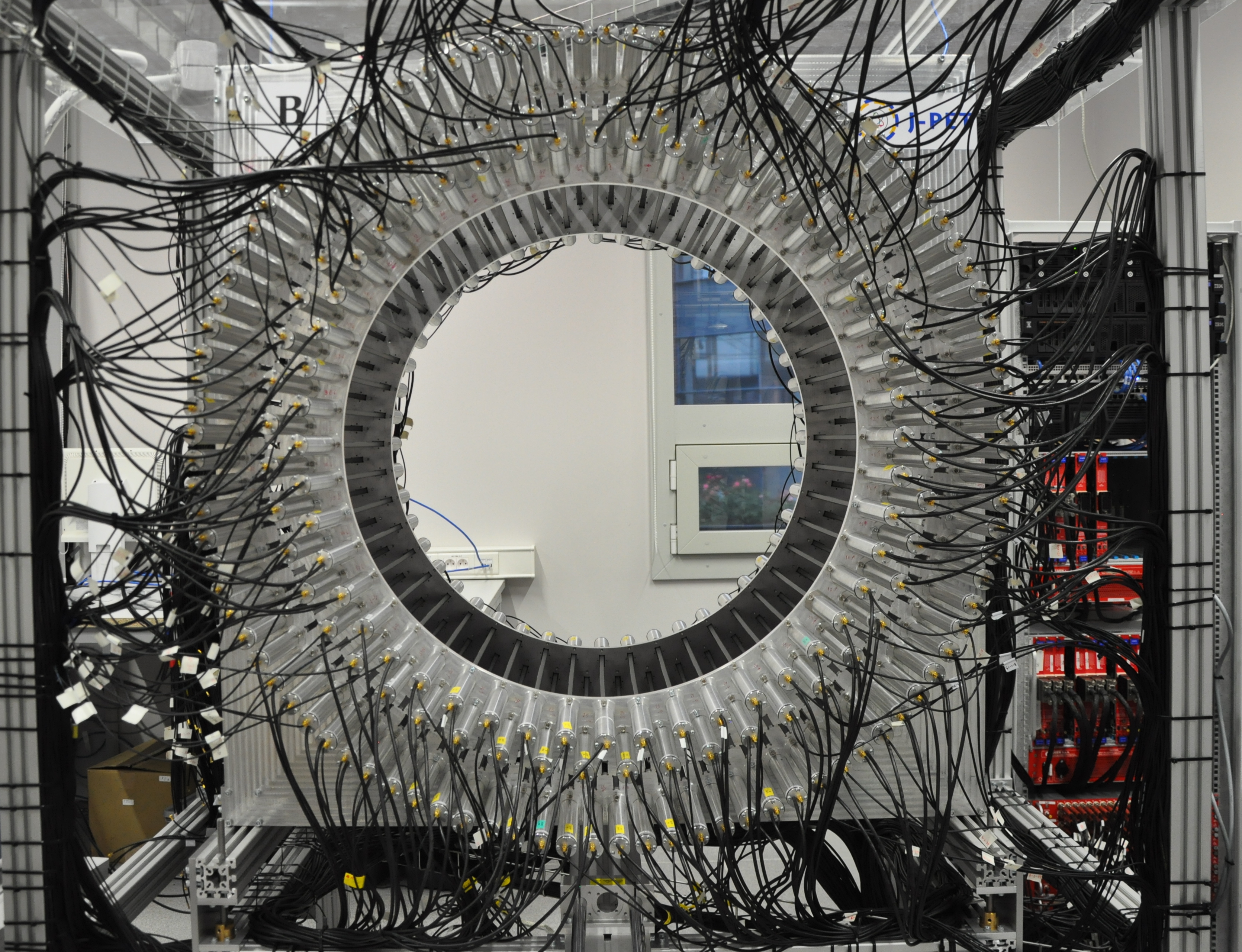}}
     \caption{\label{PhotoJPET}}
     \end{subfigure}
%    \hfill
    \begin{subfigure}{.40\textwidth}{
%    \subfigure[\label{JPETXn}]{%
    \includegraphics*[width=\textwidth]{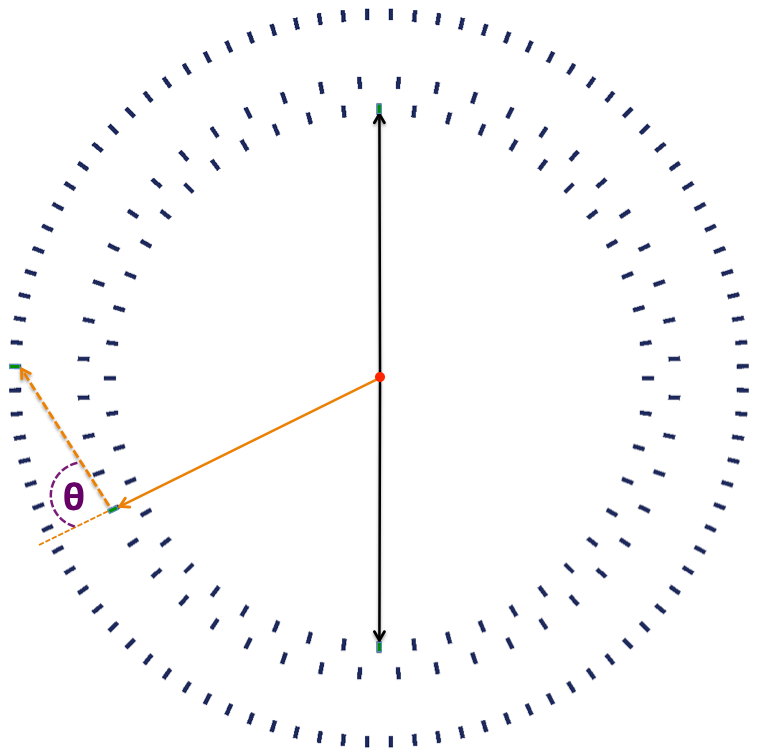}}
    \caption{\label{JPETXn}}
    \end{subfigure}
    \caption{\label{FourThreshold}(a) The photo of J-PET scanner with the front view (see text for description). (b) The cross section of J-PET with superimposed arrows depicting the interaction of 511 keV photons (black solid lines) originating from positron-electron annihilation and prompt photon 1275~keV before (orange solid line) and after (orange dashed line) scattering inside the scintillator. Red dot in the centre represents the $^{22}$Na source.}
\end{figure}

The Jagiellonian - Positron Emission Tomograph (J-PET) is constructed of 192 plastic scintillors made of EJ-230 material.~Scintillator with dimension 0.7 x 1.9 x 50 cm$^3$ connected with R9800 Hamamatsu photomultiplier on each side composes a single detection module~\cite{Niedzwiecki2017,Raczynski2015}.~The detection modules are axially arranged in three layers:~the first and second layers are composed of 48 modules whereas the third layer is the arrangement of 96 modules.~Layers are not overlaying each other and are of diameters 85 cm, 93.5~cm and 115 cm respectively.~The front view of three layer prototype of the J-PET scanner is shown in Fig.~\ref{PhotoJPET} whereas Fig.~\ref{JPETXn} exhibits its cross section.~A Multi-Voltage Threshold mezzanine (MVT) board is used to probe signals at four fixed thresholds~(80, 160, 240 and 320 mV) in voltage domain (within the accuracy of 20 ps RMS)~\cite{Palka2017} to achieve the good resolution of determining the hit time and place of interaction of photon inside the scintillator~\cite{Raczynski2014,Moskal2014,Raczynski2015,Sharma2015,Moskal2016}.
%% Figure 1
Using MVT boards, connected to readouts allow to exploit the FPGAs diffrential inputs as comparators. The signals from MVT boards are sampled using TDC implemented in the FPGA devices~\cite{JUN2018,JUN2016A,JUN2016B,Palka2014} whereas in standard approach external comparator chips are used additionally~\cite{KIM2009B,XIE2013A}. The data is stored in trigger less mode with an ability to handle the data stream with rates of about 8 Gbps~\cite{Korcyl2016b,Korcyl2018}. The position of interaction along the plastic strip is calculated based on the time difference of light signals arriving at both photomultipliers.~Signals from the plastic scintillators are very fast~(rise time $\approx$~0.5~ns, fall time $\approx$~1.8~ns)~\cite{Wieczorek2017} and are prone to much lower pile-ups with respect to crystal based detectors with order of magnitude larger fall times~\cite{Lecoq2010}.~Therefore, to avoid the dead time due to the direct charge measurement, only timing of signals is used. The photon's interaction in scintillator and the arrival time of the signals in photomultipliers at each end is measured. TOT approach is adopted instead of charge integration method. In TOT approach, the time difference between the leading and trailing edge of signal pulse crossing the applied thresholds is measured.
The schematic presentation is shown in Fig.~\ref{FourThreshold}.
%%% Figure 2 
\begin{figure}[htbp]
\centering % \begin{center}/\end{center} takes some 
\includegraphics[width=.7\textwidth]{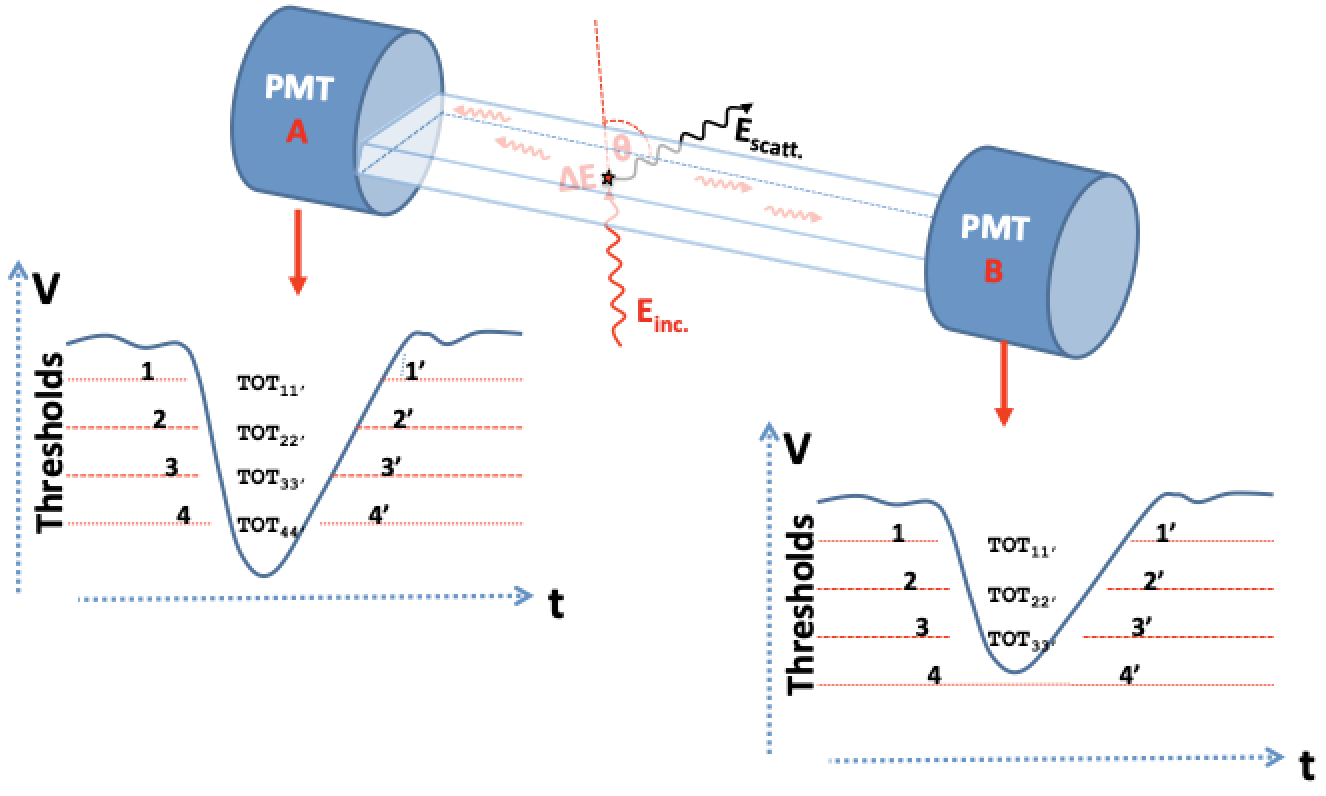}
\caption{Illustration of the analog signals obtained at photomultipliers as the outcome of the energy deposited by the photon in the interaction with plastic scintillator. Each signal is probed at four fixed thresholds. TOT value is calculated at each voltage threshold and sum of TOT's over all four thresholds applied on signals from both photomultipliers gives the resultant TOT as described by eq.~\ref{TOTdef}.}
\label{FourThreshold}
\end{figure}
The total TOT, used as a measure of the energy deposition, is estimated as a sum of TOT values measured for all thresholds at both sides of scintillator (eq.~\ref{TOTdef}):

\begin{equation}
{
\hspace{2.5cm}
TOT = \sum_{PMT = A,B} \sum_{Thr_{1-4}}^{} TOT_{PMT,Thr}\label{TOTdef}
}
\end{equation}
 Besides attractive features in terms of significant reduction in the cost of signal read out, low electricity consumption (2.76 W per channel) and high flux rate handling, the estimation of the charge collection using the pulse width of the signal is less precise and suffers a strongly non-linear relationship between TOT and energy deposition.~The plastic scintillators are composed of hydrocarbons with low atomic (Z) number, consequently the gamma photons interact in plastic scintillators predominantly through the Compton scattering and deposit only part of their energy. The information of the deposited energy is important to reduce the scatter fraction by requiring the energy loss larger than 200 keV~\cite{Moskal2014,Moskal2016,Kowalski2018A}.~To process the measured data a dedicated framework for the offline data analysis was used~\cite{Krzemien2015a,Krzemien2016}.
 %%%%%%%%%
\subsection{Experimental set-up}
\label{sec::experiment details}
\begin{figure}[htbp]
\centering % \begin{center}/\end{center} takes some 
\includegraphics[width=.75\textwidth]{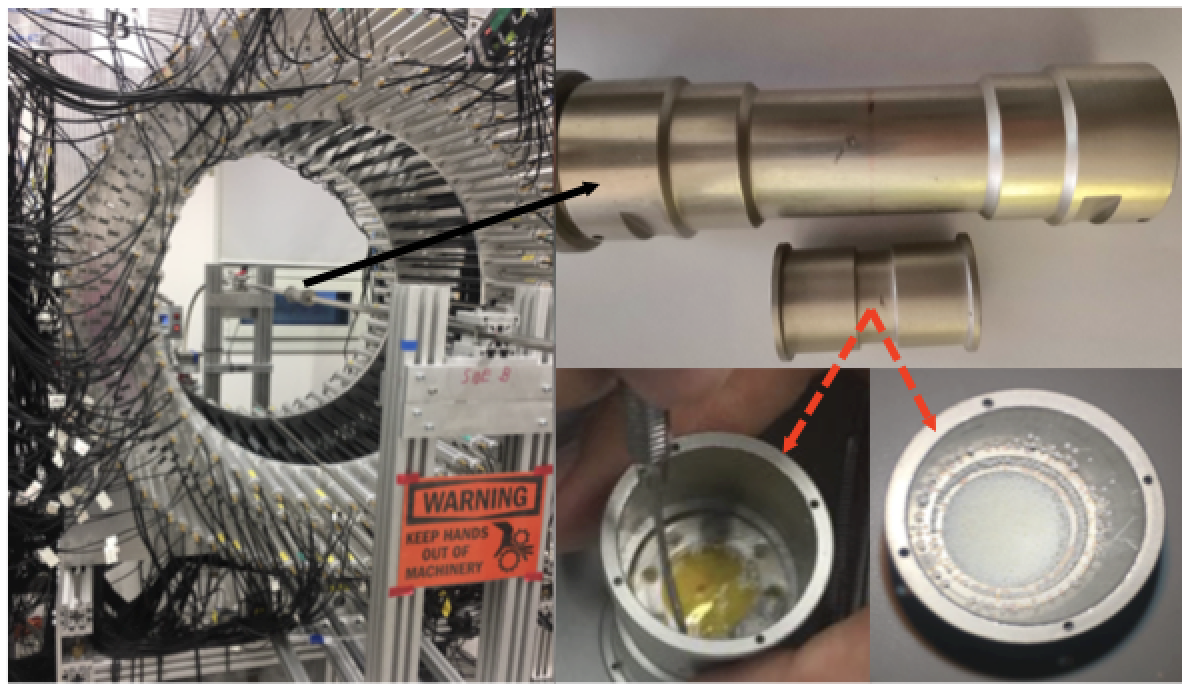}
\caption{\label{ExpSetUp} Experimental set-up of the annihilation chamber with source placed at the center of the J-PET tomograph.}
\end{figure}

The measurements were performed with a $^{22}$Na source (1 MBq activity) wrapped inside a very thin kapton foil. The experimental set-up is shown in Fig.~\ref{ExpSetUp}.~Left panel shows picture of the J-PET tomograph with the placement of a barrel shape source holder of length $\approx14$ cm and diameter $\sim~3.16$~cm~(at the center).~The source surrounded by the porous material was put inside a small chamber of thin layer made of aluminium and placed at the center of holder (see upper inset of right panel in Fig.~\ref{ExpSetUp}).~The placement of the source covered with porous material is shown at lower right insets. In this experiment, the XAD-4 polymer was used which increases the probability of formation of Ps atoms~\cite{Jasinska2016}. 
%
%%Figure 3
 Using the $^{22}$Na source gives a possibility to estimate the lifetime of the Ps atom by registering the annihilation photons and prompt gamma. 
%%%%%%%%%%%
%%% Figure 4
\begin{figure}[htbp]
    \centering
    \begin{subfigure}{.49\textwidth}{
 %   \subfloat[\label{3a}]{%
       \includegraphics*[width=\textwidth]{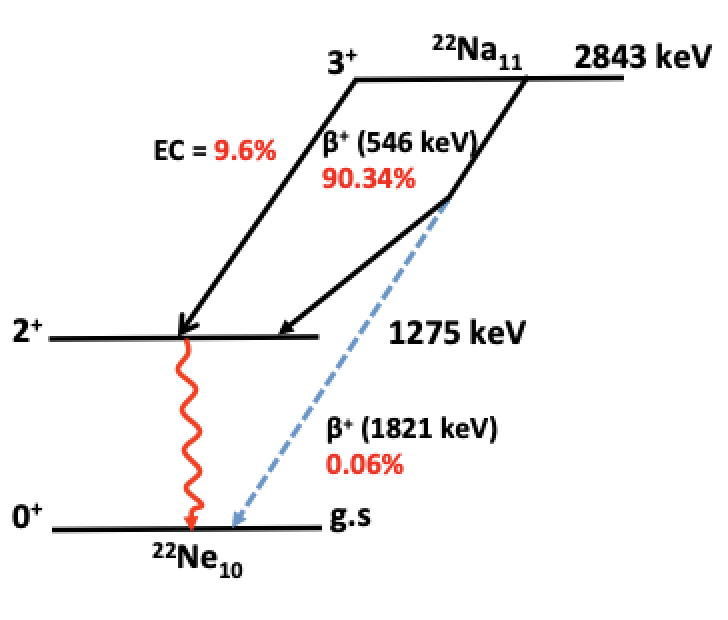}}
       \caption{\label{4a}}
       \end{subfigure}
    \hfill
    \begin{subfigure}{.45\textwidth}{
%  \subfloat[\label{3b}]{%
        \includegraphics*[width=\textwidth]{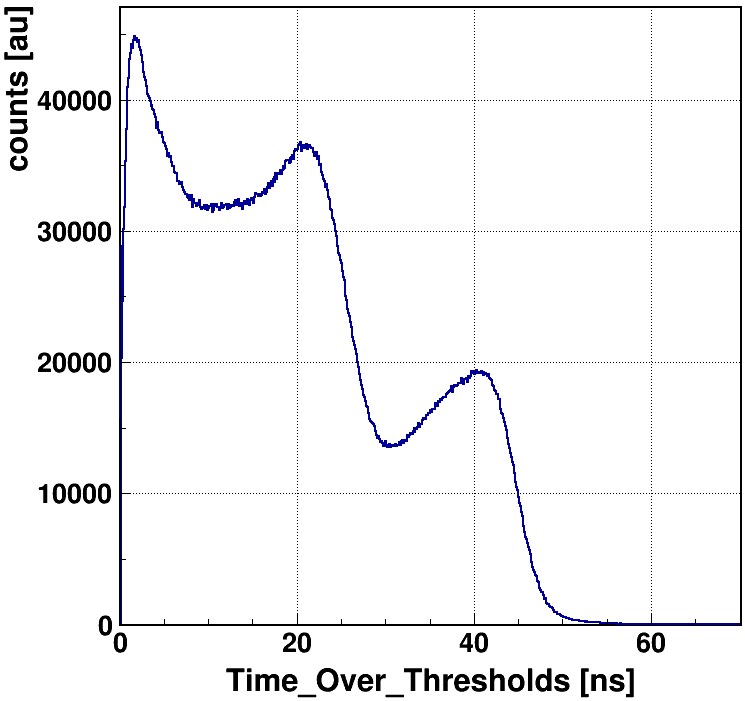}}
        \caption{\label{4b}}
        \end{subfigure}
    \caption{(a)~Above panel shows the decay scheme of the $^{22}$Na source.~(b)~Below panel shows the typical TOT spectrum obtained using the $^{22}$Na source.}
\end{figure}
 In the decay of $^{22}$Na source~(Fig.~\ref{4a}), the positron is emitted leaving behind an excited state of $^{22}$Ne nucleus which de-excite eventually via emission of prompt photon on average within the time interval of $\approx$~3.7~ps.~The time difference between the interaction of annihilation and prompt photons is then with a very good approximation equal to lifetime of Ps atom~\cite{Moskal2019}.~Gamma photons mainly interact inside the plastic scintillator via Compton scattering.\newline
 ~Fig.~\ref{4b} shows the TOT spectrum for all registered photons. The clear Compton edges are visible at about 22 ns and 46 ns corresponding to 511~keV and 1275~keV photons respectively.~The enhancement of contribution for the lower values of TOT (below 10 ns) is from the interaction of scattered photon and photons originating from the 3-photon decay of o-Ps atoms.
%%%%%
It is worth to emphasize that TOT values could be used to disentangle between the annihilation and prompt photons to some extent.~However, in the overlapping region for TOT values below 30 ns the identification would not be unambiguous. Therefore, we developed a dedicated algorithm to uniquely identify the photons of energies 511~keV and 1275~keV irrespective of TOT values (described in sec.~\ref{sec::SelectionProcedure}). 
%%% Figure 5
\begin{figure}[h]
\centering % \begin{center}/\end{center} takes some 
\includegraphics[width=.35\textwidth]{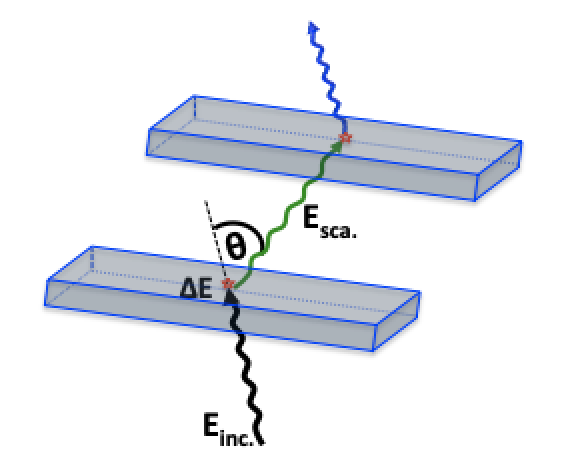}
\caption{\label{EngDepDepict} The incoming and scattered photons interactions are presented with two exemplary scintillators. Known positions of emission of $E_{inc}$ photon and positions of interactions of $E_{inc}$ and $E_{sca}$ photons inside the scintillators allow to estimate the scattering anlge.}
\end{figure}
  Furthermore, the registration of initial and scattered photons allows to determine the scattering angle $\theta$ (see Fig.~\ref{EngDepDepict}).~Known energy ($E_{inc}$) and scattering angle ($\theta$) of the initial photon give access to energy deposition $\Delta$E inside the scintillator~\cite{Compton1923}:
 \begin{equation}
 \hspace{2.5cm}
\Delta{E}=E_{inc}\left [ 1-\frac{1}{1+\frac{E_{inc}}{511}\left ( 1-cos\theta \right )} \right ]
\label{EdepVsTheta}
\end{equation}
  Determination of $\Delta$E and the measurement of TOT enables to establish the relation between these two variables.\newline
  In this work, events with three interactions were studied, assuring that all three hits forming the events are occurring in distinct scintillators. Furthermore, a photon interaction inside the scintillator is termed as hit.~The required information (E$_{inc}$,$\theta$) can be extracted based on two hits only. However, the third hit allows the event categorization in a way that we can differentiate the photons of different energies and conjecture their origins. \newline
%%%%  
\subsection{Events selection}
\label{sec::SelectionProcedure}
In this analysis, we first select the events with three hits only. The angular correlations between the registered photons in three hit events can be used partially to identify the origin and energy of photons~\cite{Kaminska2016,Czerwinski2017}. 
%%% Figure 6
\begin{figure}[h]
    \centering
    \begin{subfigure}{.44\textwidth}{
 %   \subfloat[\label{6a}]{%
       \includegraphics*[width=\textwidth]{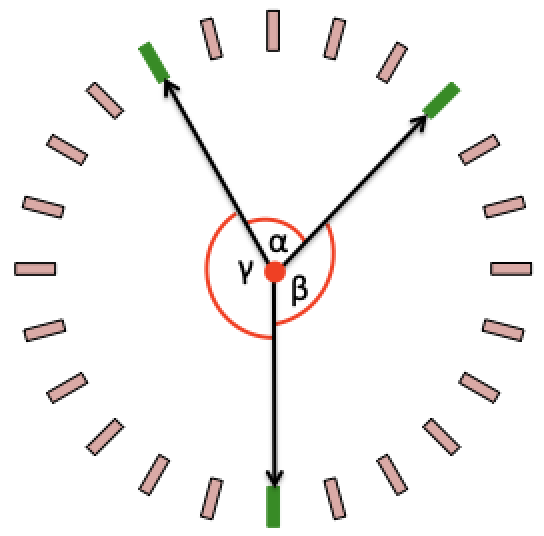}}
       \caption{\label{6a}}
       \end{subfigure}
    \hfill
    \begin{subfigure}{.44\textwidth}{
 % \subfloat[\label{6b}]{%
        \includegraphics*[width=\textwidth]{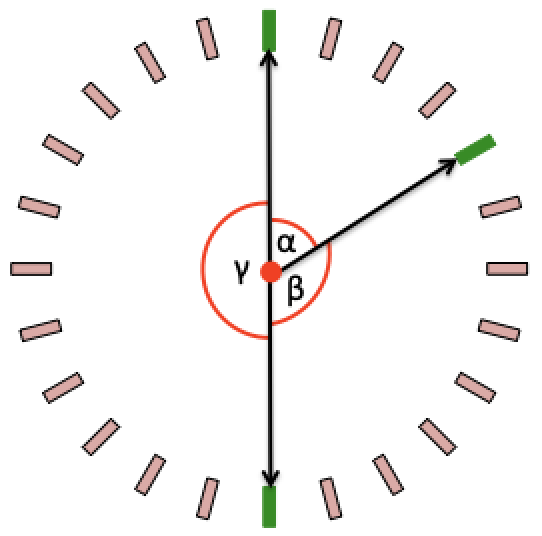}}
        \caption{\label{6b}}
        \end{subfigure}
    \hfill
    \begin{subfigure}{.44\textwidth}{
 % \subfloat[\label{6c}]{%
        \includegraphics*[width=\textwidth]{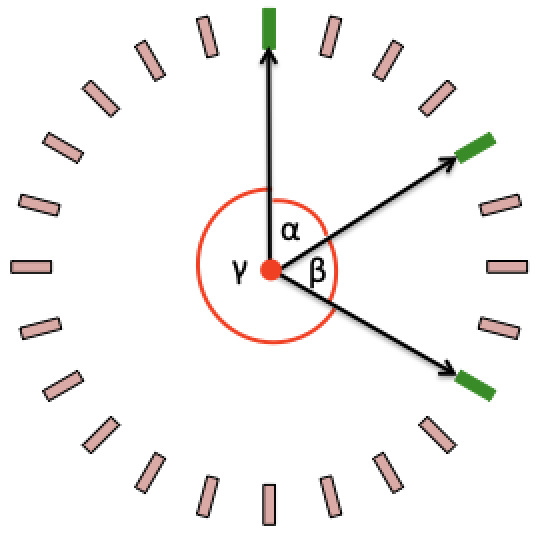}} 
        \caption{\label{6c}}
        \end{subfigure}
        \hfill
        \begin{subfigure}{.49\textwidth}{
 % \subfloat[\label{6d}]{%
        \includegraphics*[width=\textwidth]{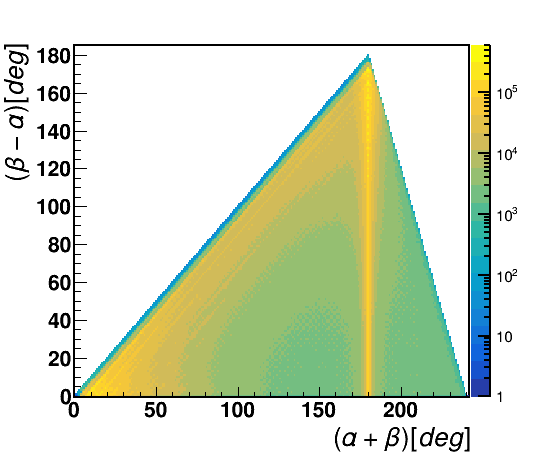}} 
        \caption{\label{6d}}
        \end{subfigure}
    \caption{(a), (b) and (c) present the three cases based on the angular correlation between the 3 hits in an event. The origin of shown events is explained in the text.~(a) correspond to the case when the sum of two smallest angles is greater than 180$^0$, (b) shows the case when sum is 180$^0$, (c) sum of two smallest angles is less than 180$^0$ and (d) the measured scatter plot of the difference ($\beta-\alpha$) vs the sum ($\alpha+\beta$) of the smallest angles}.
\end{figure}
The pictorial view of the three hits (green rectangles) and their angular correlations are presented in Fig.~\ref{6a} - \ref{6c}. For the clarity, only few scintillators mimicking the one layer of J-PET scanner are shown. The angles between three hits ordered from the smallest to the largest are $\alpha$,~$\beta$ and $\gamma$ respectively. A measured sum of the two smallest angles vs their difference is shown in Fig.~\ref{6d} which shows three distinct regions:~(i) ($\alpha$+$\beta$) $>$ 180 degree, electron-positron annihilation into three photons originating from direct $e^+e^-$ annihilation or from the decay of o-Ps atoms (ii) ($\alpha$+$\beta$) = 180, two back-to-back photons originating either from direct $e^+e^-$ annihilation or from the decay of p-Ps atom (Singlet state of Ps) or from the pick-off of o-Ps (iii) ($\alpha$+$\beta$) $<$ 180, one or two of the hits are from scattering of initial photon or prompt photon.
\subsection{Identification of 511 keV photons}
\label{sec::511keV}
\begin{figure}[htbp]
    \centering
    \begin{subfigure}{.41\textwidth}{
   % \subfloat[\label{B2BDEPICT}]{%
       \includegraphics*[width=\textwidth]{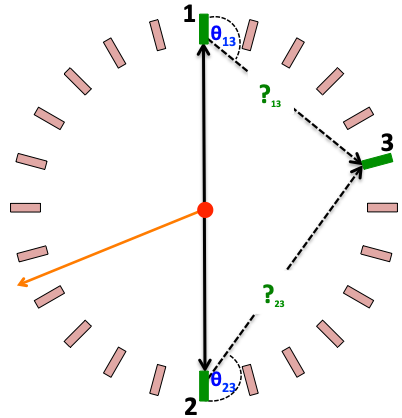}}
       \caption{\label{B2BDEPICT}}
       \end{subfigure}
    \hfill
\begin{subfigure}{.53\textwidth}{
 % \subfloat[\label{ScatterCut}]{%
        \includegraphics*[width=\textwidth]{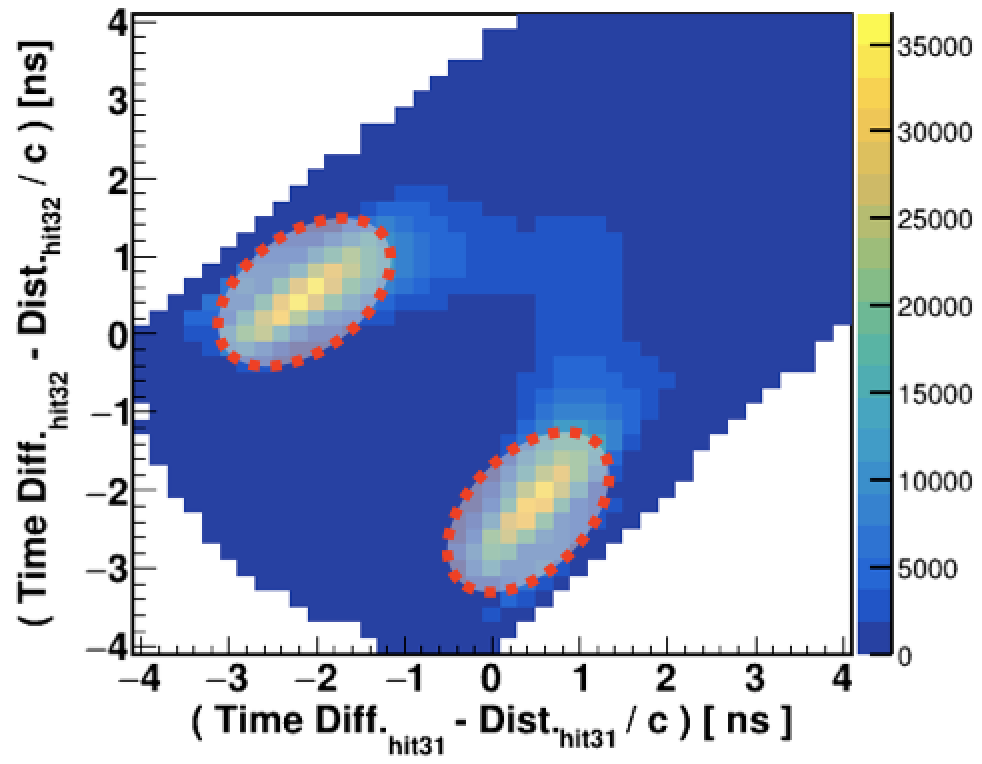}}
        \caption{\label{ScatterCut}}
        \end{subfigure}
    \caption{(a) Pictorial representation of the event selection based on the algorithm used to tag the 511 keV photons. (b) The encircled dotted red lines shows the selected scattering events of back-to-back photons based on the scatter test \textbf{S}.}
    \label{fig: 511 keV Selection}
\end{figure}

The 511 keV photons are the outcome of $e^+e^-$ annihilation into back-to-back direction to preserve the momentum conservation and can easily be identified in Fig.~\ref{6d} as a vertical line at $\alpha$+$\beta$ =180$^0$. The selection of such events was done based on the condition that the sum of the two smallest angles ($\alpha + \beta$) should be $180^0$ within the uncertainty of two degrees. The pictorial presentation of J-PET with one layer is used to visualize the situation. In Fig.~\ref{B2BDEPICT}, black lines represent the back-to-back photons labeled as 1 and 2 and orange line shows the prompt photon.   
%
%%%% Figure 7 
Scintillators with photons hits are shown as green reactangles. Hit~3 corresponds to the interaction of the scattered photon after Hit~1 or Hit~2. The exact assignment of scattering photons to its primary one is important for the proper estimation of scattering angle (Fig.~\ref{ScatterCut}) which is essential for exact energy deposition.
For this purpose, a scatter test \textbf{S} is devised which estimates the time difference between the measured time (time difference between two hits) and calculated time (distance between hit-positions divided by the speed of light). The \textbf{S}-test was applied event-wise assuming both possibility that the third hit might belong to the scattering photon after Hit~1 or Hit~2. The results are shown in Fig.~\ref{ScatterCut}. In case of the proper assignment the value of $\textbf{S}$ should be close to zero. For the final analysis, events from the areas encircled by red-dotted lines were taken into account.\newpage
With the known incident photon's energy (511 keV) and its scattering angle, the deposited energy is calculated in response to the initial hit. Thus, a one-to-one information of energy deposition by 511 keV photons and corresponding TOT values on event-wise bases is extracted. In studying 511~keV photons the relationship can be established only upto 340~keV energy deposition (Compton edge for 511 keV photon). So, In order to extend the relationship for higher energy deposition values, the studies with the prompt photons with energy 1275 keV are also performed. The selection criterion for the prompt gamma is explained in the next sub-section.  
\subsection{Identification of 1275 keV photons and their scatterings}
The selection of prompt and its corresponding scatter photon without using the explicit cut on TOT values is not as straight forward as in the case of annihilation photons. 
%
%%%%% Figure 8
\begin{figure}[htbp]
    \centering
    \begin{subfigure}{.31\textwidth}{
   % \subfloat[\label{PromptpPs}]{%
       \includegraphics*[width=\textwidth]{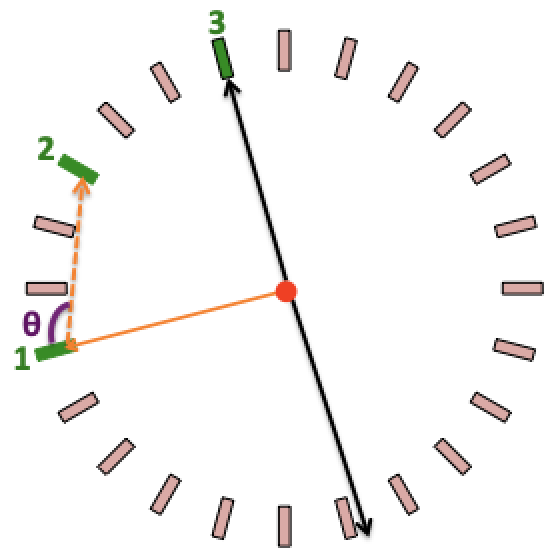}}
       \caption{\label{PromptpPs}}
       \end{subfigure}
    \hfill
    \begin{subfigure}{.31\textwidth}{
  %\subfloat[\label{PromptoPs}]{%
        \includegraphics*[width=\textwidth]{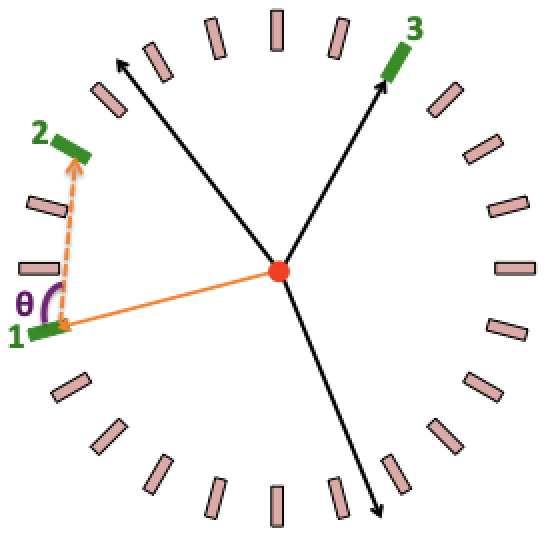}}
        \caption{\label{PromptoPs}}
        \end{subfigure}
 %       \vspace{-0.8cm}
    \hfill
\begin{subfigure}{.31\textwidth}{
  %\subfloat[\label{PromptBG}]{%
        \includegraphics*[width=\textwidth]{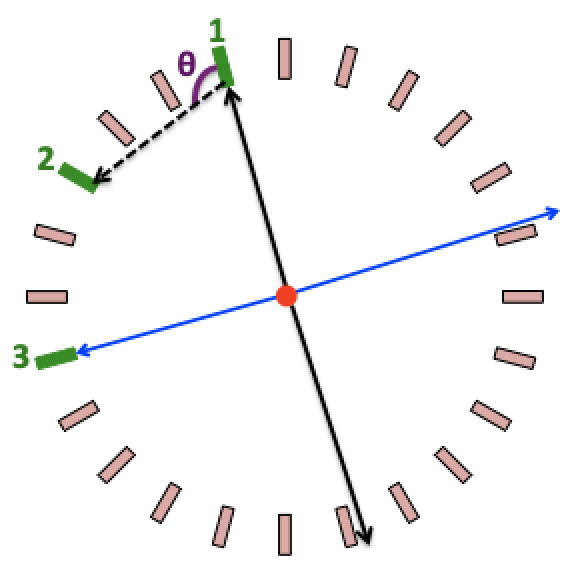}}
        \caption{\label{PromptBG}}
        \end{subfigure}
    \caption{ Three different cases in the events selection for the 1275 keV photons (orange arrows) based on the delayed decay of ortho-Positronium atoms: (a) first case shows the annihilation photons from the decay of para-Positronium (or ortho-Positronium pick-off), (b) shows the photons originating from the decay of ortho-Positronium state and ~(c) presents the possible source of background in the adopted procedure for the selection of prompt photon due to the accidental coincidences. The black and blue solid lines presenting the back-to-back annihilation photons might originate from two different decays.}
    \label{deexcitation_gamma}
\end{figure}
As the first step in the selection procedure, in order to suppress events with back-to-back 511 photons, we applied the cut based on the angular correlation between the hits (Fig.~\ref{6d}) such that we are not considering those events when the sum of two smallest angles lie in interval 165$^0$~-~185$^0$ degrees.
In the next step, hits are ordered according to the ascending time and only those events are selected for which time interval between first and third hit is larger than 10 ns. The assumed analogy is as follows:~Hit 1 is considered to be the prompt photon, Hit 2 is the scattered photon of the prompt, and Hit 3 is assumed by one of the photons originating from pick-off~(Fig.~\ref{PromptpPs}) or direct from the decay of ortho-positronium atoms~(Fig.~\ref{PromptoPs})
%%
%%%% Figure 9
\begin{figure}[htbp]
\centering 
\includegraphics*[width=.60\textwidth]{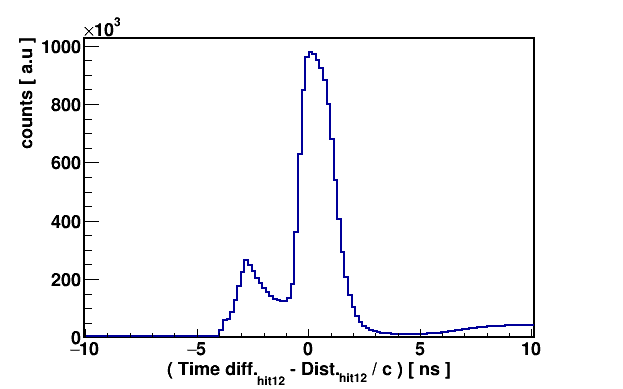}
\caption{\label{deexct_cut} Results of \textbf{S}-test calculated between Hit~1 and Hit~2 in event-wise manner. For the true scattering of prompt photons the value of S-test is chosen in between -0.5 to 0.75 ns.}
\end{figure}
Fig.~\ref{PromptBG} shows the possible background in the adopted selection criterion.~The time difference ($\Delta t$) between the first and third hit is used as a key parameter. It is chosen to be much larger than the possible time difference between the scatterings of photon in the tomograph. For the present analysis, the value for $\Delta$t was fixed in time interval (10~ns - 100~ns).
For estimating the scattering angles of prompt gamma, the scattered test was applied between first and second hit.~The histogram used for the scatter test is shown in Fig. \ref{deexct_cut}. The acceptance time window for the true scattering of the $\textbf{S}$-test is chosen between -0.5 to 0.75~ns.   
\section{Results}
\label{sec::results}

\begin{figure}[htbp]
    \centering
    \begin{subfigure}{.78\textwidth}{
    %\subfloat[\label{TOTVsEdep}]{%
    \includegraphics[width=\textwidth]{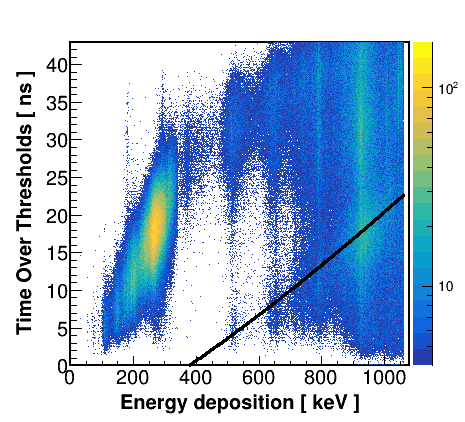}}
    \caption{\label{TOTVsEdep}}
    \end{subfigure}
    \hfill
    \centering
    \begin{subfigure}{.72\textwidth}{
 % \subfloat[\label{TOTVsEdep2}]{%
        \includegraphics[width=\textwidth]{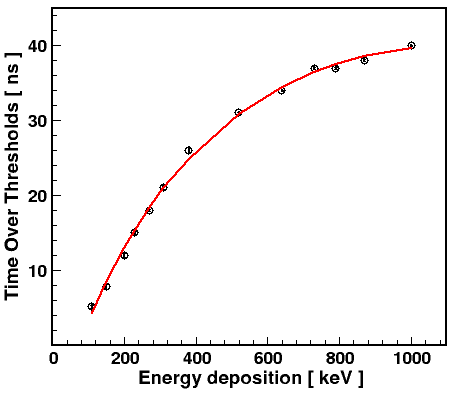}}
        \caption{\label{TOTVsEdep2}}
        \end{subfigure}
    \caption{ (a) 2-D spectrum of TOT versus energy deposition. (b) TOT vs energy deposition. Red line indicates result of the fit of the function: TOT = A0 + A1 * ln(E$_{dep}$ + A2) + A3 * (ln(E$_{dep}$ + A2))$^2$, with A0 = -2332 ns, A1 = 632.1 ns/keV, A2 = 606.9 keV and A3 = -42.08 ns/(keV)$^2$.  
}
    \label{deexcitation_gamma}
\end{figure}

The motivation of this work is the elaboration of a method to estimate the energy loss by incident photons in the interaction with plastic scintillators used in J-PET tomography scanner.
The analysis is performed with the $^{22}$Na isotope because it is long lived~(2.6 years half lifetime) and is emitting positrons annihilating to 511~keV photons as used in the PET diagnostics.
Since, the aim of this study is to establish the relation between TOT and $\triangle$E for the plastic scintillators, in order to avoid the biasness in identification of 511~keV (positron annihilation) and 1275~keV (deexcitation) photons, a new method of photon identification was developed which is independent of the TOT values. After selecting the photons of different energies, their scattering angles are estimated based on the hit characteristics. With the information of incident energy of photon and its scattering direction, the energy transfer to the electrons inside the scintillator can be calculated using the Compton scattering formula (eq.~\ref{EdepVsTheta}). \\
In each event, hit characteristics are measured that comprise the TOT values, photon's interaction time as well as spatial coordinates.
For all selected hits, the relationship between measured TOT values and the estimated energy deposition is presented in Fig.~\ref{TOTVsEdep}.~An enhancement in TOT values is observed with increasing energy depositions which is an expected behavior. However, some contribution is visible with large energy deposition but less TOT values, which can be interpreted as the wrongly tagged photon due to the accidental coincidences i.e., instead of 1275 keV photon, photons with lower incident energies were used to estimate the energy deposition for the scattering angles.(e.g., see the Fig. \ref{PromptBG}). The black line in~\ref{TOTVsEdep} is plotted by assuming that the wrongly tagged photons are of energy 511 keV.
%
%%% Figure 10
%
%%% TOT vs Eng dep -
%
For the final relationship, the profile histograms of Fig.~\ref{TOTVsEdep} for the most populated energy bins are selected. The mean values of the TOT distributions as a function of the center value of the energy interval is shown in Fig.~\ref{TOTVsEdep2}.
The black circles are the experimental data. Errors in estimating the mean value of the TOT are within the size of the used symbols. It is shown that a simple function (see the caption of Fig.~\ref{deexcitation_gamma}) is able to reproduce the data for quite a large range of energy deposition , i.e. up to 950 keV.
\section{Conclusions}
\label{sec::Conclusions}
The J-PET is first PET scanner composed of plastic scintillators~\cite{Moskal2016,Niedzwiecki2017}.~Plastic scintillators are more than an order of magnitude less expensive than the crystal scintillators.~Time-Over-Threshold approach facilitates a compact, fast and cost effective signal readout~\cite{Palka2017,Korcyl2016b,Korcyl2018}. In the framework of J-PET scanner, TOT approach is adopted as the measure of the energy deposition to utilize the fast timing and low pile-up features of the plastics scintillators.~To use this method effectively in comparison to the classical method of charge collection, it is important to know the non-linear relationship between the energy deposition of an initial photon and the measured TOT values. Here, we presented  a method for determining relation between the energy deposition and TOT which can be efficiently applied to the J-PET tomograph by collecting the data with the $^{22}$Na source covered with the porous material characterized with the long lifetime of positronium atoms~\cite{Jasinska2016}. The identification of an incident photon was based on angular correlation between the three hits and the lifetime of the metastable positronium atoms while the scattered photon was identified and associated to the initial photon based on the correlation among the time and distance between the interaction points. Using $^{22}$Na source emitting 1275 keV prompt and 511 keV annihilation photons, the TOT versus energy loss relationship up to about 1000 keV was established. The proposed function fits well the experimental data and can be used as a standalone function for the energy loss calibration of the plastic scintillators used in the J-PET scanner.\\

\begin{backmatter}
\section*{Ethics approval and consent to participate}
Not applicable
\section*{Consent for publication}
Not applicable
\section*{Availability of data and material}
The data that support the findings of this study are available from the corresponding author upon
 request.
\section*{Competing interests}
The authors declare that they have no competing interests
\section*{Funding}
This work was supported by The Polish National Center for Research and Development through grant INNOTECH-K1/IN1/64/159174/NCBR/12, the Foundation for Polish Science through the MPD and TEAM PIOR.04.04.00-00-4204/17, the National Science Centre of Poland through grants no. 2016/21/B/ST2/01222, 2017/25/N/NZ1/00861, the Ministry for Science and Higher Education through grants no. 6673/IA/SP/2016, 7150/E- 338/SPUB/2017/1, 7150/E-338/M/2017, 7150/E-338/M/2018 and the Austrian Science Fund FWF-P26783-N27.
\section*{Authors' contributions}
All authors contributed in the elaboration of methods for data analysis, interpretation of results, experimental set-up and data collection, correction of manuscript and approval. The analysis algorithm was mainly developed by SS along with the PM. SS and PM edited the manuscript.
\section*{Acknowledgements}
The authors acknowledge technical and administrative support of A.~Heczko, M.~Kajetanowicz and W.~Migda\l{}. 
\bibliographystyle{vancouver}
\bibliography{TOT_vs_Edep}

\begin{thebibliography}{10}

\bibitem{Nygren1991}
Nygren DR.
\newblock Converting vice to virtue: can time-walk be used as a measure of
  deposited charge in silicon detectors?
\newblock Internal LBL note. May, 1991;.

\bibitem{Kipnis1997}
Kipnis I, Collins T, DeWitt J, Dow S, Frey A, Grillo A, et~al.
\newblock {A Time-over-threshold machine: The readout integrated circuit for
  the BABAR Silicon Vertex Tracker}.
\newblock IEEE Transactions on Nuclear Science. 1997;44(3 PART 1):289--297.

\bibitem{Moskal2016}
Moskal P, Rundel O, Alfs D, Bednarski T, Bia{\l}as P, Czerwi{\'{n}}ski E,
  et~al.
\newblock {Time resolution of the plastic scintillator strips with matrix
  photomultiplier readout for J-PET tomograph}.
\newblock Physics in Medicine and Biology. 2016;61(5):2025--2047.

\bibitem{Moskal2018}
Moskal P, Krawczyk N, Hiesmayr BC, Bala M, Curceanu C, Czerwinski E, et~al.
\newblock {Feasibility studies of the polarization of photons beyond the
  optical wavelength regime with the J-PET detector}.
\newblock Eur Phys J C. 2018;78:970.

\bibitem{SAINTGOBAIN}
Saint Gobain Crystals.
\newblock http://wwwcrystalssaint-gobaincom;.

\bibitem{Wieczorek2017}
Wieczorek A, Dulski AK, Niedzwiecki S, Alfs D, Bialas P, Curceanu C, et~al.
\newblock {Novel scintillating material 2-(4-styrylphenyl)benzoxazole for the
  fully digital and MRI compatible J-PET tomograph based on plastic
  scintillators}.
\newblock PLos One. 2017;12(11):e0186728.

\bibitem{CIMA2006}
Vilardi I, Braem A, Chesi E, Ciocia F, Colonna N, Corsi F, et~al.
\newblock {Optimization of the effective light attenuation length of YAP:Ce and
  LYSO:Ce crystals for a novel geometrical PET concept}.
\newblock Nucl Instr $\&$ Meth A. 2006;564:506--514.

\bibitem{RMAO}
Mao R, Zhang L, Ren-Yuan Z.
\newblock {Optical and scintillation properties of inorganic scintillators in
  high energy physics}.
\newblock IEEE Transactions on Nuclear Science. 2008;55:2425--2431.

\bibitem{LABR3}
Lanthanum Bromide Properties.
\newblock
  https://wwwcrystalssaint-gobaincom/products/standard-and-enhanced-lanthanum-bromide;.

\bibitem{Olcott2008}
Olcott PD, Levin CS.
\newblock {Pulse width modulation: A novel readout scheme for high energy
  photon detection}.
\newblock IEEE Nuclear Science Symposium Conference Record.
  2008;(c):4530--4535.

\bibitem{fujiwara2010}
Fujiwara T, Takahashi H, Shimazoe K, Shi B.
\newblock {Multi-Level Time-Over-Threshold Method for Energy Resolving
  Multi-Channel Systems}. 2010;57(5):2545--2548.

\bibitem{Orita2018}
Orita T, Koyama A, Yoshino M, Kamada K, Yoshikawa A, Shimazoe K, et~al.
\newblock {The current mode Time-over-Threshold ASIC for a MPPC module in a
  TOF-PET system}.
\newblock Nuclear Instruments and Methods in Physics Research, Section A:
  Accelerators, Spectrometers, Detectors and Associated Equipment.
  2018;912:303--308.

\bibitem{Kim2009}
Kim H, Kao CM, Xie Q, Chen CT, Zhou L, Tang F, et~al.
\newblock {A multi-threshold sampling method for TOF-PET signal processing}.
\newblock Nuclear Instruments and Methods in Physics Research, Section A:
  Accelerators, Spectrometers, Detectors and Associated Equipment.
  2009;602(2):618--621.

\bibitem{Grant2014}
Grant AM, Levin CS.
\newblock {A new dual threshold time-over-threshold circuit for fast timing in
  PET}.
\newblock Physics in Medicine and Biology. 2014;59(13):3421--3430.

\bibitem{Shimazoe2012}
Shimazoe K, Takahashi H, Shi B, Orita T, Furumiya T, Ooi J, et~al.
\newblock {Dynamic time over threshold method}.
\newblock IEEE Transactions on Nuclear Science. 2012;59(6):3213--3217.

\bibitem{Orita2015}
Orita T, Shimazoe K, Takahashi H.
\newblock {The dynamic time-over-threshold method for multi-channel APD based
  gamma-ray detectors}.
\newblock Nuclear Instruments and Methods in Physics Research, Section A:
  Accelerators, Spectrometers, Detectors and Associated Equipment.
  2015;775:154--161.

\bibitem{VAN2016}
Vandenberghe S, Mikhaylova E, D'Hoe E, Mollet P, Karp JS.
\newblock {Recent developments in time-of-flight PET}.
\newblock EJNMMI Physics. 2016;3:1--30.

\bibitem{Slomka2016}
Slomka PJ, Pan T, Germano G.
\newblock {Recent Advances and Future Progress in PET Instrumentation}.
\newblock Seminars in Nuclear Medicine. 2016;46:5--19.

\bibitem{Jin-Jie2008}
Jin-Jie W, Yue-Kun H, Zhi-Jia S, Chong W, Yu-Da Z, Gui-An Y, et~al.
\newblock {A study of time over threshold (TOT) technique for plastic
  scintillator counter}.
\newblock Chinese Physics C. 2008;32(3):186--190.

\bibitem{Ashrafi2011}
Ashrafi S, Ghahremani~Gol M.
\newblock {Energy calibration of thin plastic scintillators using Compton
  scattered $\gamma$rays}.
\newblock Nuclear Instruments and Methods in Physics Research, Section A:
  Accelerators, Spectrometers, Detectors and Associated Equipment.
  2011;642(1):70--74.

\bibitem{Kowalski2016A}
Kowalski P, W{\"{i}}licki W, Raczy{\'{n}}ski L, Alfs D, Bednarski T, Bia{\l}as P,
et.al.
\newblock {Scatter fraction of the J-PET tomography scanner}.
\newblock Acta Physica Polonica B. 2016;47:549--560.

\bibitem{Kowalski2018A}
Kowalski P, W{\"{i}}licki W, Shopa RY, Raczy{\'{n}}ski L, Klimaszewski K,
  Curceanu C, et~al.
\newblock Estimating the NEMA characteristics of the J-PET tomograph using the
  GATE package.
\newblock Physics in Medicine and Biology. 2018;63(16):165008--165025.

\bibitem{Compton1923}
Compton HA.
\newblock The spectrum of scatterd x-rays.
\newblock {Physical Review}. 1923;22(20):2--6.

\bibitem{Garwin1953}
Garwin RL.
\newblock {Thermalization of positrons in metals}.
\newblock Phys Rev. 1953;91:1571--1572.

\bibitem{Moskal2019}
Moskal P, Kisielewska D, Curceanu C, Czerwi{\'{n}}ski E, Dulski K, Gajos A,
  et~al.
\newblock {Feasibility study of the positronium imaging with the J-{PET}
  tomograph}.
\newblock Phys Med Biol. 2019;64(5):055017.

\bibitem{Moskal2019NatureReview}
Moskal P, Jasińska B, Stępień E, Bass S.
\newblock Positronium in Medicine and Biology.
\newblock Nature Reviews Physics. 2019;1:527--529.

\bibitem{Dulski2018}
Dulski K, Curceanu C, Czerwi{\'{n}}ski E, Gajos A.
\newblock {Commissioning of the J-PET detector in view of the positron
  annihilation lifetime spectroscopy}.
\newblock Hyperfine Interactions. 2018;239(40):1--6.

\bibitem{Jasinska2016}
Jasinska B, Gorgol M, Wiertel M, Zaleski R, Alfs D, Bednarski T, et~al.
\newblock {Determination of the 3 $\gamma$ fraction from positron annihilation
  in mesoporous materials for symmetry violation experiment with J-PET
  scanner}. 2016;47(2):453--460.

\bibitem{Niedzwiecki2017}
Nied{\'{z}}wiecki S, Bia{\l}as P, Curceanu C, Czerwi{\'{n}}ski E, Dulski K,
  Gajos A, et~al.
\newblock {J-PET: A New Technology for the Whole-body PET Imaging}.
\newblock Acta Physica Polonica B. 2017;48(10):1567--1576.

\bibitem{Raczynski2015}
Raczy{\'{n}}ski L, Moskal P, Kowalski P, Wi{\'{s}}licki W, Bednarski T, Bia P,
  et~al.
\newblock {Compressive sensing of signals generated in plastic scintillators in
  a novel J-PET instrument}.
\newblock Nuclear Instruments and Methods in Physics Research, Section A:
  Accelerators, Spectrometers, Detectors and Associated Equipment.
  2015;786:105--112.

\bibitem{Palka2017}
Pa{\l}ka M, Strzempek P, Korcyl G, Bednarski T, Nied{\'{z}}wiecki S, Bia{\l}as
  P, et~al.
\newblock {Multichannel FPGA based MVT system for high precision time (20 ps
  RMS) and charge measurement}.
\newblock Journal of Instrumentation. 2017;12:P08001.

\bibitem{Raczynski2014}
Raczy{\'{n}}ski L, Moskal P, Kowalski P, Wi{\'{s}}licki W, Bednarski T,
  Bia{\l}as P, et~al.
\newblock {Novel method for hit-position reconstruction using voltage signals
  in plastic scintillators and its application to Positron Emission
  Tomography}.
\newblock Nuclear Instruments and Methods in Physics Research, Section A:
  Accelerators, Spectrometers, Detectors and Associated Equipment.
  2014;764:186--192.

\bibitem{Moskal2014}
Moskal P, Nied{\'{z}}wiecki S, Bednarski T, Czerwi{\'{n}}ski E, Kap{\l}on,
  Kubicz E, et~al.
\newblock {Test of a single module of the J-PET scanner based on plastic
  scintillators}.
\newblock Nuclear Instruments and Methods in Physics Research, Section A:
  Accelerators, Spectrometers, Detectors and Associated Equipment.
  2014;764:317--321.

\bibitem{Sharma2015}
Sharma NG, Silarski M, Bednarski T, Bialas P, Czerwi{\'{n}}ski E, Gajos A,
  et~al.
\newblock {Reconstruction of hit time and hit position of annihilation quanta
  in the J-PET detector using the Mahalanobis distance}.
\newblock Nukleonika. 2015;60(4):765--769.

\bibitem{JUN2018}
Won JY, Lee JS.
\newblock Highly Integrated FPGA-only Signal Digitization Method Using
  Single-ended Memory Interface Input Receivers for Time-of-Flight PET
  Detectors.
\newblock IEEE Transactions on Biomedical Circuits and Systems.
  2018;2865581:1--10.

\bibitem{JUN2016A}
Won JY, Kwon SI, Yoon HS, Ko GB, Son JW, Lee JS.
\newblock Dual-Phase Tapped-Delay-Line Time-to-Digital Converter With
  On-the-Fly Calibration Implemented in 40 nm FPGA.
\newblock IEEE Transactions on Biomedical Circuits and Systems.
  2016;10:231--242.

\bibitem{JUN2016B}
Won JY, Lee JS.
\newblock Time-to-Digital Converter Using a Tuned-Delay Line Evaluated in 28-,
  40-, and 45-nm FPGAs.
\newblock IEEE Transactions on Instrumentation and Measurement.
  2016;65:1678--1689.

\bibitem{Palka2014}
Pa{\l}ka M, Moskal P, Bednarski T, Bia{\l}as P, Czerwi{\'{n}}ski E, Kap{\l}on
  {\L}, et~al.
\newblock {A novel method based solely on field programmable gate array (FPGA)
  units enabling measurement of time and charge of analog signals in positron
  emission tomography (PET)}.
\newblock Bio-Algorithms and Med-Systems. 2014;10:41--45.

\bibitem{KIM2009B}
Kim H, Kao CM, Xie Q, Chen CT, Zhou L, Tang F, et~al.
\newblock A multi-threshold sampling method for TOF-PET signal processing.
\newblock Nucl Instr and Meth A. 2009;602:618–621.

\bibitem{XIE2013A}
Xi D, Zheng C, Liu W, Liu X, Wan L, Kim H, et~al.
\newblock A PET detector module using FPGA-only MVT digitizers.
\newblock Nuclear Science Symposium and Medical Imaging Conference (NSS/MIC)
  IEEE. 2013;10:1--5.

\bibitem{Korcyl2016b}
Korcyl G, Alfs D, Bednarski T, Bia{\l}as P, Czerwi{\'{n}}ski E, Dulski K,
  et~al.
\newblock {Sampling FEE and Trigger-less DAQ for the J-PET Scanner}.
\newblock Acta Physica Polonica B. 2016;47(2):491--496.

\bibitem{Korcyl2018}
Korcyl G, Bia{\l}as P, Curceanu C, Czerwi E, Hiesmayr BC, Jasi B, et~al.
\newblock {Evaluation of Single-Chip , Real-Time Tomographic Data Processing on
  FPGA SoC Devices}.
\newblock IEEE Transactions on Medical Imaging. 2018;37(11):2526--2535.

\bibitem{Lecoq2010}
Lecoq P, Auffray E, Brunner S, hILLEMANNS H, Jarron P, Knapitsch A, et~al.
\newblock {Factors Influencing Time Resolution of Scintillators and Ways to
  Improve Them}.
\newblock IEEE Transactions on Nuclear Science. 2010;57(5):2411--2416.

\bibitem{Krzemien2015a}
Krzemie{\'{n}} W, Bala M, Bednarski T, Bialas P, Czerwi{\'{n}}ski E, Gajos A,
  et~al.
\newblock {Processing optimization with parallel computing for the J-PET
  scanner}.
\newblock Nukleonika. 2015;60(4):745--748.

\bibitem{Krzemien2016}
Krzemie{\'{n}} W, Alfs D, Bia{\l}as P, Czerwi{\'{n}}ski E, Gajos A, G{\l}owacz
  B, et~al.
\newblock {Overview of the software architecture and data flow for the J-PET
  tomography device}.
\newblock Acta Physica Polonica B. 2016;47(2):561--567.

\bibitem{Kaminska2016}
Kami{\'{n}}ska D, Gajos A, Czerwi{\'{n}}ski E, Alfs D, Bednarski T, Bia{\l}as
  P, et~al.
\newblock {A feasibility study of ortho-positronium decays measurement with the
  J-PET scanner based on plastic scintillators}.
\newblock European Physical Journal C. 2016;76(445):1--14.

\bibitem{Czerwinski2017}
Czerwi{\'{n}}ski E, Dulski K, , Bia{\l}as P, Curceanu C, Czerwi{\'{n}}ski E,
  et~al.
\newblock {Commissioning of the J-PET detector for studies of decays of
  positronium atoms}.
\newblock Acta Physica Polonica B. 2017;48(10):1961--1968.

\end{thebibliography}
%%%%% Figures in sequence %%%%%%%%%
% Figure 1

%
% Figure 2

%
%\newpage
% Figure 3
%
%%% Figure 4 

%
%%%%%% Figure 5
%
%%%% Figure 6
%%%%
%%%% Figure 7 

%%%%%
%%% Figure 8
%%%%%%%
%%% Figure 9
%%%%
%%%%%% Figure 10
%%%%%
%\bibliographystyle{vancouver}
%\bibliography{TOT_vs_Edep}
\end{backmatter}
\end{document}